\documentclass[iop]{emulateapj}
\slugcomment{{\sc Accepted to AJ:} September 26, 2014}

\renewcommand\plotone[1]{\includegraphics[width=\linewidth]{#1}}
\newcommand\masyear{mas~year$^{-1}$}
                                                 
\shorttitle{Deep Proper Motion Catalog}
\shortauthors{Munn et al.}

\begin{document}

\title{A Deep Proper Motion Catalog Within The Sloan Digital Sky Survey
Footprint}

\author{Jeffrey A. Munn and Hugh C. Harris}
\affil{US Naval Observatory, Flagstaff Station, 10391 W. Naval Observatory
Road, Flagstaff, AZ 86005-8521; jam@nofs.navy.mil, hch@nofs.navy.mil}

\author{Ted von Hippel}
\affil{Embry-Riddle Aeronautical University, Physical Sciences,
600 S. Clyde Morris Blvd, Daytona Beach, FL 32114-3900; ted.vonhippel@erau.edu}

\author{Mukremin Kilic}
\affil{University of Oklahoma, Homer L. Dodge Department of Physics and
Astronomy, 440 W. Brooks Street, Norman, OK 73019; kilic@ou.edu}

\author{James W. Liebert}
\affil{University of Arizona, Steward Observatory, Tucson, AZ 85721;
jamesliebert@gmail.com}

\author{Kurtis A. Williams}
\affil{Department of Physics and Astronomy, Texas A\&M University--Commerce,
P.O. Box 3011, Commerce, TX 75429; kurtis.williams@tamuc.edu}

\author{Steven DeGenarro}
\affil{Department of Astronomy, University of Texas at Austin,
  1 University Station C1400, Austin, TX 78712-0259; studiofortytwo@yahoo.com}

\author{Elizabeth Jeffery}
\affil{BYU Department of Physics and Astronomy, N283 ESC, Provo, UT 84602;
ejeffery@byu.edu}

\and

\author{Trudy M. Tilleman}
\affil{US Naval Observatory, Flagstaff Station, 10391 W. Naval Observatory
Road, Flagstaff, AZ 86005-8521; trudy@nofs.navy.mil}

\begin{abstract}
A new proper motion catalog is presented, combining the Sloan Digital
Sky Survey (SDSS) with second epoch observations in the $r$ band
within a portion of the SDSS imaging footprint.  The new observations
were obtained with the 90prime camera on the Steward Observatory Bok
90 inch telescope, and the Array Camera on the U.S. Naval Observatory,
Flagstaff Station, 1.3 meter telescope.  The catalog covers 1098
square degrees to $r = 22.0$, an additional 1521 square degrees to $r
= 20.9$, plus a further 488 square degrees of lesser quality data.
Statistical errors in the proper motions range from 5~\masyear\ at the
bright end to 15~\masyear\ at the faint end, for a typical epoch
difference of 6 years.  Systematic errors are estimated to be roughly
1~\masyear\ for the Array Camera data, and as much as 2 -- 4~\masyear\ for the
90prime data (though typically less).  The catalog also includes a second epoch
of $r$ band photometry.

\end{abstract}


\section{Introduction}

Wide area, deep proper motion catalogs have wide applicability to
studies of the structure and formation of our Galaxy.  Most current
catalogs are based on photographic surveys done with the various 1
meter-class Schmidt telescopes, including the Palomar Oschin 1.2 meter
(POSS-I and POSS-II), the European Southern Observatory 1.0 meter (ESO
surveys), and the United Kingdom 1.2 meter Schmidt Telescope (SERC and
AAO surveys).  Recent proper motion catalogs based on these Schmidt
plates include USNO-B \citep{monet2003} and the LSPM high proper motion
catalog \citep{lspm}.  These catalogs are limited by the depth of the
Schmidt plates to $V \sim 20$.

The Sloan Digital Sky Survey
\citep[SDSS;][]{york2000,gunn1998,gunn2006,fukugita1996} is the modern
successor to the Schmidt surveys.  Using CCDs rather than photographic
plates, SDSS provides 5-band ($ugriz$) imaging over 14,555 square
degrees, with a 95\% completeness rate for point sources of $r =
22.2$, thus reaching roughly 2 magnitudes fainter than the Schmidt
surveys.  However, SDSS is primarily a single epoch survey, and thus proper
motions cannot be derived from SDSS data alone.  While a number of
studies have combined SDSS with USNO-B to produce improved proper
motions \citep[e.g.,][]{munn2004,gould2004}, these are still limited to the
depth of the Schmidt plates.

This paper presents new observations targeting 3100 square degrees in
the SDSS footprint, separated in epoch from the SDSS observations by
typically 5 -- 10 years.  These are combined with SDSS astrometry to
create a new proper motion catalog, extending 1 -- 2 magnitudes
fainter than the Schmidt based catalogs.  The immediate science driver
for the new catalog is to extend the work of \citet{harris2006} on the
white dwarf luminosity function, with the goal of greatly increasing
samples of the coolest disk white dwarfs as well as thick disk and
halo white dwarfs (see \citealt{kilic2010} for three halo white dwarfs
discovered in this survey).  Moreover, the catalog should be broadly useful
for a range of Galactic and stellar research.

\section{Observations}

The initial survey was conducted using the 90prime prime focus wide-field
imager on the Steward Observatory Bok 90 inch telescope \citep{williams2004}.
90prime is a mosaic of four Lockheed 4k by 4k CCDs,
with a $1.16\arcdeg \times 1.16\arcdeg$
edge-to-edge field-of-view (FOV), an imaging area of 1.06 square degrees,
inter-CCD gaps in both dimensions of $0.131\arcdeg$, and a plate scale of
0.45 arcsec per pixel.  2107 observations were obtained over 45 nights from
2006 to 2008.  Each target field was observed once in the SDSS $r$ filter
(except for the night UT  2006 January 2, when the Bessell $R$ filter was
accidentally used).  Exposure times were 5 minutes in good conditions, and up
to 15 minutes in cloudy weather or bad seeing.  Field centers were separated
by one degree in both right ascension and declination.

There were a number of complications with the 90prime instrument at
the time of the observations, all of which have since been largely corrected.
The focal plane was populated with experimental CCDs, which have since been
replaced.  There
was a large charge trap on CCD 3, rendering data in the region $1560 <
x < 2700$ and $y > 2530$ unusable.  One amp on CCD 2, covering half of
the chip ($x > 2048$), displays a pattern noise, evident as low level
striping parallel to the x (declination) axis.  While the astrometry
and photometry are usually only minimally affected, for the nights of 2007
May 25 -- 26 and June 11 -- 12, the pattern noise was bad
enough that data on that half of the chip was unusable.  There were some
pointing issues, 
causing some observations to not completely cover the intended target
field.  Finally, the Bok telescope required refocusing and recollimation any
time the
telescope was slewed through large angles.  Evidence for instability
in the focus and/or collimation, and its effects on the astrometry, will be
discussed below.

Starting in 2009, observations were obtained using the Array Camera on
the U.S. Naval Observatory, Flagstaff Station, 1.3 meter telescope.
The Array Camera is a 2 x 3 mosaic of 2048 x 4102 e2v CCDs with
0.6$\arcsec$ pixels, and a field of view of $1.41\arcdeg$ in right
ascension (with inter-CCD gaps of $0.062\arcdeg$) by $1.05\arcdeg$ in
declination (with inter-CCD gaps of $0.019\arcdeg$).  2532
observations were obtained over 179 nights from 2009 -- 2011, using 20
minute exposure times.  Field centers were separated by 1.4 degrees in
right ascension and 1.0 degrees in declination.  As with the Bok,
there were some pointing issues with the 1.3m at the time which caused
some observations to not completely cover their intended target field.

\begin{figure}
\plotone{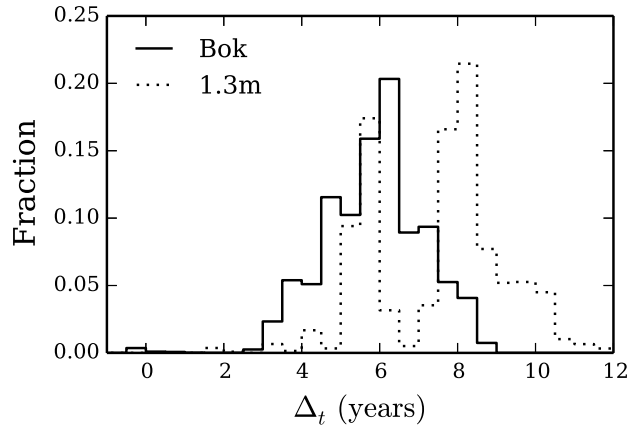}
\caption{Distribution of epoch differences between the SDSS observations
  and this survey's observations, for representative samples of stars
  in the Bok (solid histogram) and 1.3m (dotted histogram) surveys.
  }
\label{fig-epoch}
\end{figure}

Target fields were chosen so as to maximize the epoch difference between
the SDSS observations and this survey's observations, and thus minimize the
proper motion errors. Each survey field is covered by more than one SDSS scan,
and thus stars within a given field will have different epoch differences.
Figure~\ref{fig-epoch} plots the distribution of epoch differences for a
representative sample of stars, separately for the Bok and 1.3m surveys.
75\% of the Bok survey, and 97\% of the 1.3m survey, have epoch differences
greater than 5 years.

\section{Data Processing}

Each observation generates four (for 90prime) or six (for Array Camera) images,
one image for each CCD in the camera.  Each image is processed separately,
throughout all steps in the data processing.

\subsection{Object Detection and Characterization}

Images were bias subtracted and flat-field corrected using the Image
Reduction and Analysis Facility \citep[IRAF;][]{tody1986,
  tody1993}\footnote{IRAF is distributed by the National Optical
  Astronomy Observatories, which are operated by the Association of
  Universities for Research in Astronomy, Inc., under cooperative
  agreement with the National Science Foundation.}.  Median object
flats were used to flat-field correct the Bok images, while median
twilight flats were used for the 1.3m images.  SExtractor
\citep{bertin1996} was then used to detect and characterize the
objects in the images.  Object detection used a detection threshold of
1.5 times the background rms, after convolution with a Gaussian with a
FWHM of 3 pixels.  Object detection was also performed using DAOPHOT
II \citep{stetson1987}, which does a better job detecting stars near
overexposed stars.  The merged SExtractor and DAOPHOT detections were
then fed to DAOPHOT, which was used to model the point-spread function
(PSF), and to measure PSF magnitudes based on that model PSF.  The PSF
was allowed to vary quadratically across each image.

\subsection{Astrometric Calibration}

The SDSS Seventh Data Release
\citep[DR7;][]{sdssDR7} was used to provide calibration stars to both
astrometrically and photometrically calibrate the observations, as
well as to provide the first epoch positions for the proper motions.
At the time of this work, both SDSS DR8 \citep{sdssDR8} and DR9
\citep{sdssDR9} were also available.  DR8 introduced some bugs in the
astrometric processing, leading to less accurate astrometry for some
objects (this has been corrected in subsequent data releases).  DR9
uses a different algorithm from prior releases to choose the primary
detection of an object which has been observed more than once.  The
new algorithm has the effect that, in some regions of sky, DR9 uses
more recent observations than those used in DR7, leading to a shorter
epoch difference between the SDSS primary observation and our
observation, and thus less accurate derived proper motions.  DR7
represents the best choice to maximize both the quality of the
astrometry and the epoch difference with our observations.

For use as calibration stars, the SDSS positions must be propagated to
the epoch of our observations.  This is done using the proper motions
from the SDSS+USNO-B catalog \citep{munn2004,munn2008}, which
recalibrates the individual plate detections in USNO-B using SDSS
galaxies, and then combines those recalibrated positions with the SDSS
positions to derive improved proper motions.  The random errors for
the SDSS+USNO-B proper motions vary from about 2.8 -- 4.7~\masyear\
over the magnitude range $17 < r < 20$ from which
calibration stars are selected, while the systematic errors are
typically of order 0.2~\masyear, though they can be 2 -- 3
times larger in some patches of sky \citep{bond2010}.

The astrometric calibrations use the DAOPHOT centers measured by
fitting the PSF to each star.  An astrometrically clean set of
calibration stars is selected, by requiring that (1) they pass the set
of criteria suggested on the SDSS DR7 Web site for defining a clean
sample of point sources\footnote{See
  http://www.sdss.org/dr7/products/catalogs/flags.html}; (2) their $r$
magnitude be in the range 17 -- 20, so as to avoid saturated stars and
stars with large centering errors; (3) their $r-i$ color be in the
range -0.5 -- 2, for which differential chromatic refraction (DCR) can
be well-modelled as a linear function of $r-i$; and (4) their
SDSS+USNO-B proper motions be well measured, following the
prescription given in \citet{kilic2006}.  There are a minimum of 100
calibrating stars per image, and more typically 200 -- 400.  The
tangent plane coordinates of the calibration stars, $\xi$ and $\eta$,
are fit as functions of image coordinate $x$ and $y$, using the
following formulae:
\begin{eqnarray}
x' & = & x + (c_0 + c_1 \tan z \cos q) (r-i) + m_0(x,y),\\
y' & = & y + (c_0 + c_1 \tan z \sin q) (r-i) + m_1(x,y),\\
\xi & = & a_0 + a_1 x' + a_2 y'  +  a_3 x'^2 + a_4 x'y' + a_5 y'^2\\
\nonumber
& &+ a_6 x'^3 + a_7 x'y'^2 + a_8 x'^2y' + a_9 y'^3,\\
\eta & = & b_0 + b_1 x' + b_2 y'  +  b_3 x'^2 + b_4 x'y' + b_5 y'^2\\
\nonumber
& &+ b_6 x'^3 + b_7 x'y'^2 + b_8 x'^2y' + b_9 y'^3.
\end{eqnarray}
Equations (1) and (2) correct the raw image coordinates for effects
which are stable across many runs, including DCR and high order optical
distortions.  In equations (1) and (2), $x$ and $y$ are the
image coordinates of the star, $x'$ and $y'$ are the corrected
coordinates, $z$ is the zenith distance, $q$ is the parallactic angle,
$r$ and $i$ are the SDSS magnitudes, $c_i$ are the corrections for
DCR, and $m_i$ are residual maps used to correct high order
optical distortions.  These corrected image coordinates
are then fit to the calibration star tangent plane coordinates
$\xi$ and $\eta$ in equations (3) and (4), using cubic polynomials.

\begin{figure}
\plotone{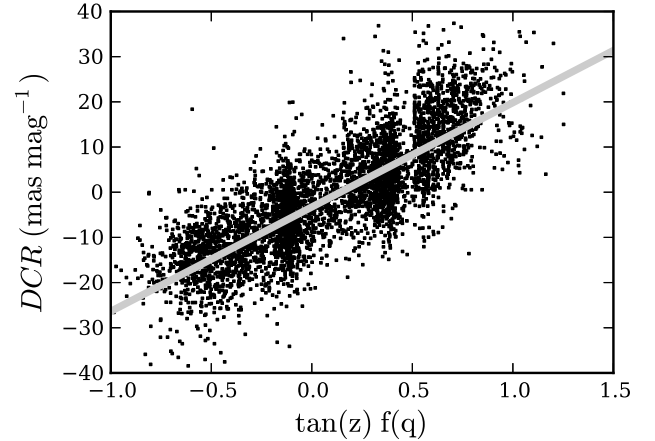}
\caption{Differential chromatic refraction (DCR), measured along both
  the x and y axes, for all images in the 1.3m survey, plotted against
  a factor proportional to the expected total refraction along that
  axis (the total refraction is proportional to the tangent of the
  zenith distance, $z$, and $f(q)$ accounts for the component of DCR
  along each axis, where $q$ is the parallactic angle, and $f(q)$ is
  $\rm{cos}(q)$ for the the x-axis and $\rm{sin}(q)$ for the y-axis).
}
\label{fig-dcr}
\end{figure}

To derive the correction terms $c_i$ and $m_i$ in equations (1) and
(2), equations (3) and (4) are first solved for each image, using
uncorrected image coordinates.  For each image, DCR is then solved for
by fitting the residuals in $x$ and $y$, separately, as linear
functions of $r - i$.  Since this is a noisy measurement for
individual frames, we then fit the ensemble of the measured DCR for
all images (separately for each survey) against the expected total
atmospheric refraction along the axis (which is proportional to the
tangent of the zenith distance, with the component along each axis
determined by the parallactic angle).  The results for the 1.3m are
shown in Figure~\ref{fig-dcr} (the results for the Bok are similar).
This fit determines the values of $c_o$ and $c_1$ in equations (1) and
(2).  DCR corrections are fairly small in the SDSS $r$ filter,
amounting to less than 20 mas~mag$^{-1}$ for both surveys at a zenith
distance of 45 degrees.  There is measurable DCR even at the zenith,
where no atmospheric DCR is to be expected, of 4.4 mas~mag$^{-1}$ for
the 90prime and 3.3 mas~mag$^{-1}$ for the 1.3m; this is not currently
understood, but we note that both cameras feature refractive field correctors.

\begin{figure*}
\plotone{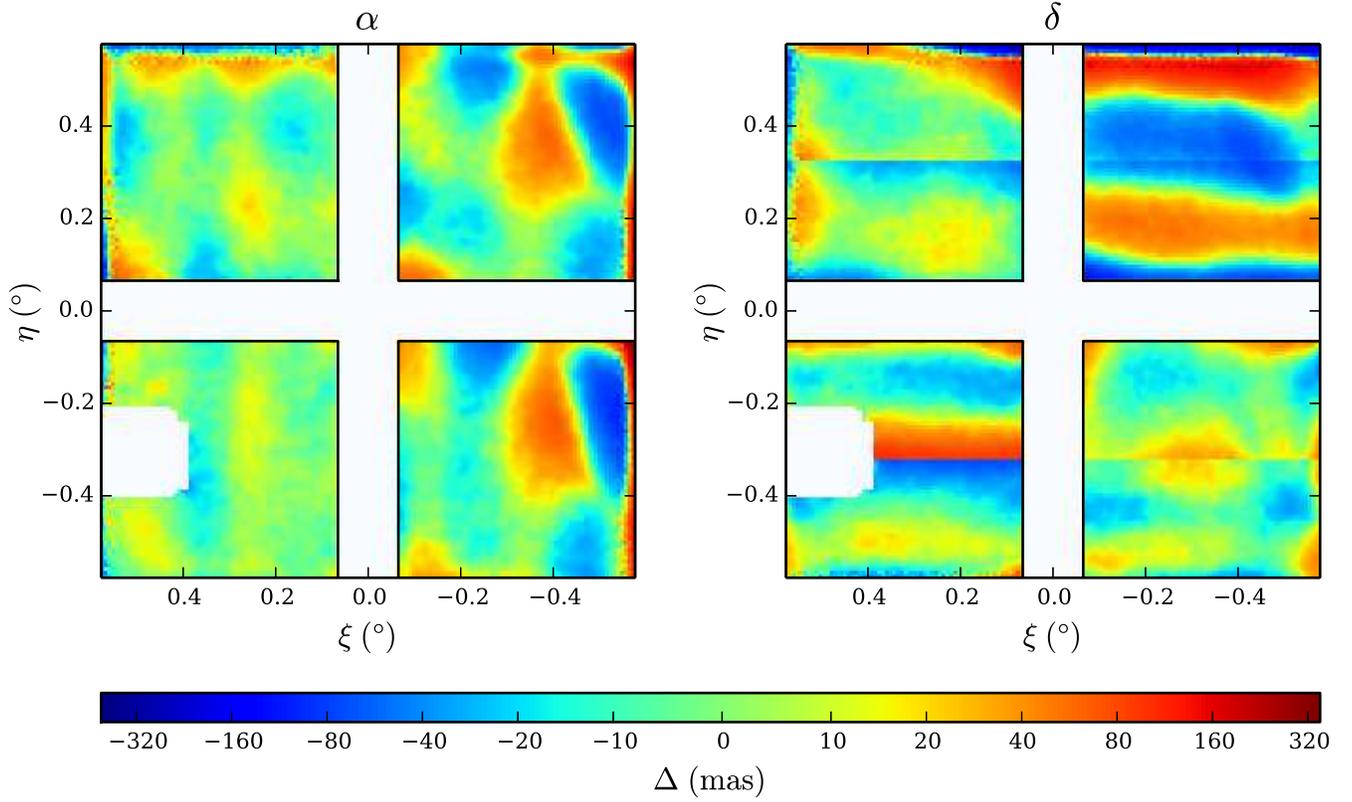}
\caption{Map showing the mean residuals (catalog - observed, in mas)
  across the focal plane for the 2008 Bok runs.  Residuals in right
  ascension are shown in the left figure, and in declination in the
  right figure.}
\label{fig-bok-residual-map}
\end{figure*}

\begin{figure*}
\plotone{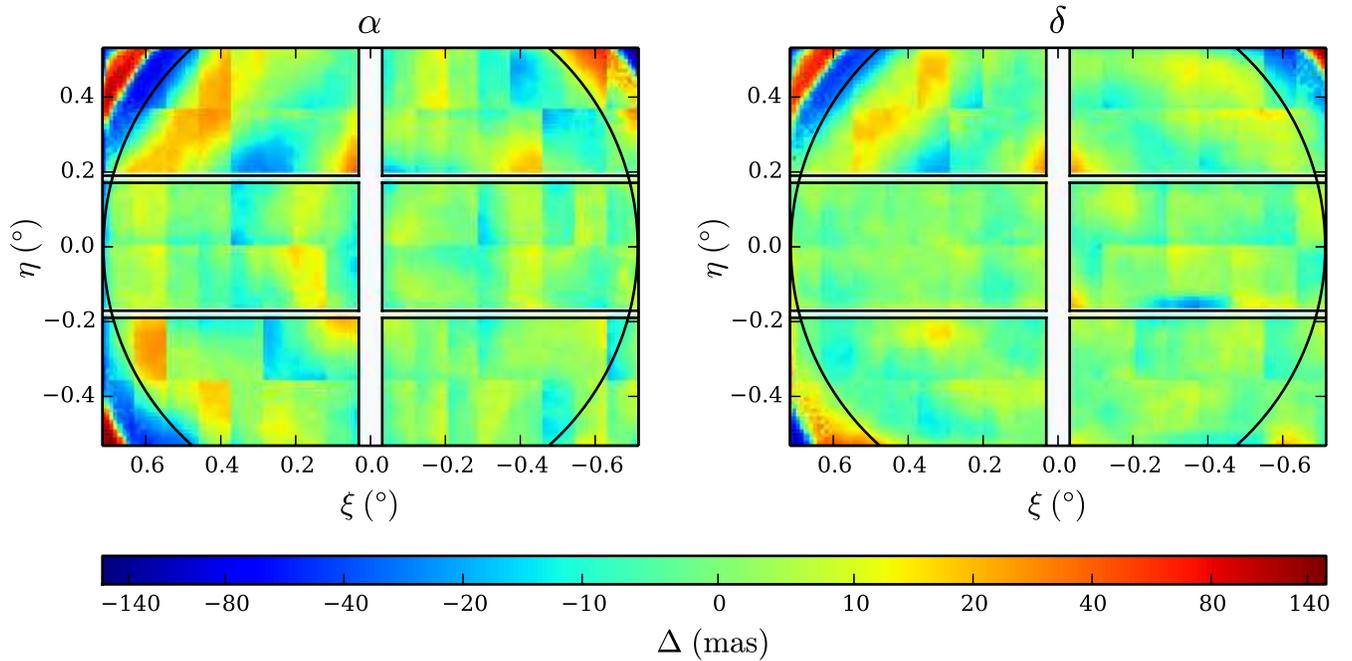}
\caption{Map showing the mean residuals (catalog - observed, in mas)
  across the focal plane for the 1.3m survey.  Residuals in right
  ascension are shown in the left figure, and in declination in the
  right figure.  A circle with radius 0.7 degrees is shown for reference.}
\label{fig-13m-residual-map}
\end{figure*}

After applying the DCR corrections, residual maps were created by measuring
the mean residuals across multiple nights, binned across the focal plane.
For the 1.3m, a single residual map was used for the entire survey, except
for one dark run comprising nine nights of observations from 2009 April 19
through May 1, for which there were clear remaining systematic residuals
on CCD 4, and thus a different residual map was used for that CCD only.  We
suspect there was a problem with the primary mirror support system during
that run.  For the Bok, separate residual maps were used
for the 2006 and 2007 runs, and the 2008 runs.
Figures~\ref{fig-bok-residual-map} and \ref{fig-13m-residual-map} display
the residual maps for the Bok (2008 runs) and 1.3m, respectively.  These
residual maps are then used to correct the raw image coordinates for
high order optical distortions (the $m_i$ terms in equations (1) and (2)).

Finally, the plate solutions for the individual images (equations (3)
and (4)) are rederived, now using coordinates corrected for DCR and
high order optical distortions (equations (1) and (2)).  Proper
motions are then calculated by simply differencing our positions and the
positions of the matching SDSS DR7 primary detections.

The stability of the high order optical distortions was monitored by
examining residual maps averaged over individual observing runs and,
for the Bok survey, individual nights (due to the longer exposure
times for the 1.3m, there aren't enough observations in individual
nights to generate statistically significant residual maps).  For the
Bok survey, variable large scale systematic errors across the focal
plane are seen, of up to 10 --- 20 mas (though typically considerably
less over most of the FOV), both from observing run to observing run,
as well as night to night within observing runs.  The greatest
variability is seen for the declination residuals on CCDs 2 and 4.
While much of the data are better than this, these variable residuals
represent a source of irreducible field-dependent systematic error.
These can introduce systematic errors in the proper motions of up to
4~\masyear\ (though more typically half that), dependent on position
in the focal plane, and varying in time.  There are not enough
calibration stars on individual exposures to remove these
time-dependent variations in the residual maps.  The 1.3m shows no
such run-to-run variations, with the exception of the single dark run
discussed above.  However, again due to the considerably smaller
number of exposures taken per night on the 1.3m, it is not possible to
look for night-to-night variations.

\begin{figure}
\plotone{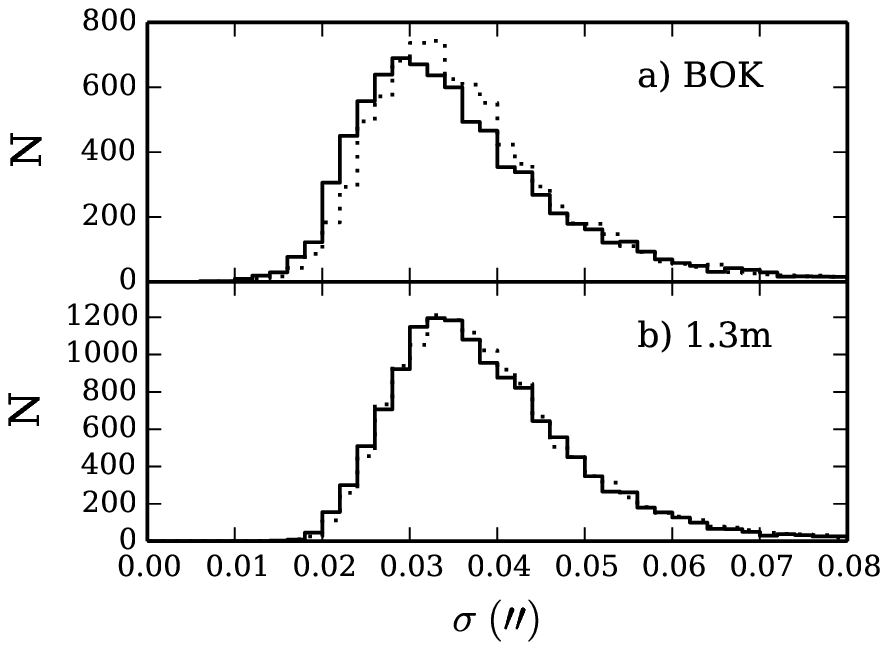}
\caption{Distribution of astrometric calibration errors for each
  image, for the Bok (panel (a)) and 1.3m (panel (b)) surveys.  The
  distributions in right ascension (solid histograms) and declination
  (dotted histograms) are displayed separately.}
\label{fig-calib-rms}
\end{figure}

The errors in the calibrations may be characterized by the rms
residuals in the final fits for each image, limited to bright unsaturated
stars ($17 < r < 18$ for the Bok survey, $16.5 < r < 18$ for the 1.3m survey)
for which centering errors make negligble contributions.
Figure~\ref{fig-calib-rms} displays histograms of the rms residuals for
each image, separately for each survey, and separately in right ascension and
declination.  The distributions peak at 30 and 35 mas for the Bok and 1.3m
surveys, respectively.

\subsection{Photometric Calibration}

DAOPHOT PSF magnitudes were calibrated directly against SDSS PSF magnitudes,
using sets of SDSS calibration stars similar to those used for the astrometric
calibrations, over the magnitude range $17 < r < 20$.  
The PSF varies considerably over the large FOVs of both telescopes, thus the
calibration is dependent on the stars' positions in the FOV.
The PSF magnitudes were first corrected for color terms, using a separate
correction for each survey, linear with $r-i$, and constant over each survey.
For each image separately, a two-dimensional cubic surface was fit to the
differences between the SDSS PSF magnitudes and color-corrected DAOPHOT
magnitudes, using the formula
\begin{eqnarray}
r_{sdss} - r'_{psf} & = & a_0 + a_1 x + a_2 y + a_3 x^2 + a_4 xy + \\
\nonumber
& & a_5 y^2 + a_6 x^3 + a_7 xy^2 + a_8 x^2y + a_9 y^3,
\end{eqnarray}
where $r_{sdss}$ is the SDSS PSF magnitude, $r'_{psf}$ is the
color-corrected DAOPHOT PSF magnitude, and $x$ and $y$ are the image
coordinates of the star.  There are a minimum of 200 calibrating stars
per image, and more typically 400 -- 600.

\begin{figure}
\plotone{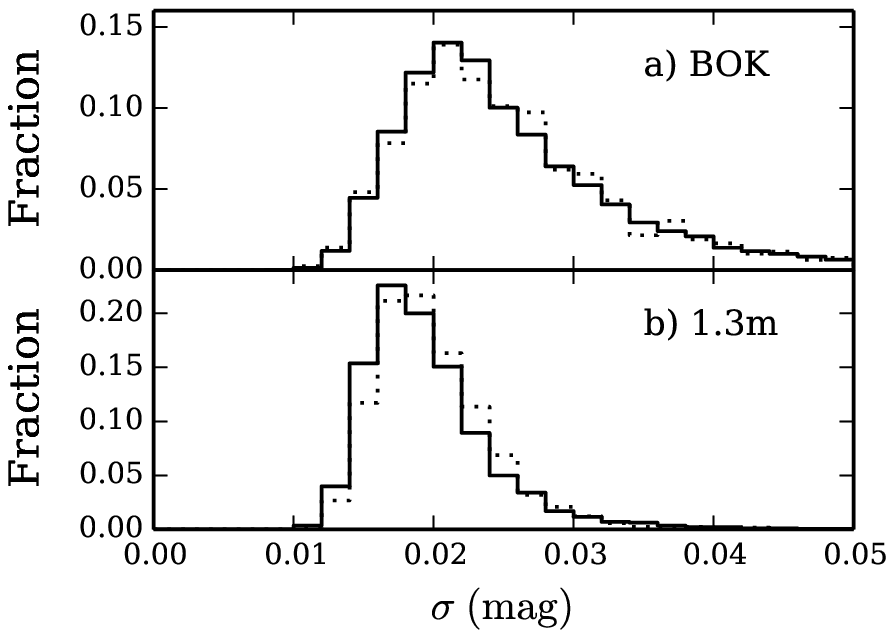}
\caption{Distribution of photometric calibration errors for each
  image, for the Bok (panel (a)) and 1.3m (panel (b)) surveys.
    Solid and dotted histograms are the distributions for
  ``good'' and ``ok'' fields, respectively.}
\label{fig-pcalib-rms}
\end{figure}

The calibration errors may be estimated using the rms residuals for the brighter
calibration stars ($r < 18$).  Figure~\ref{fig-pcalib-rms}
displays histograms for the estimated calibration errors.  Separate
histograms are displayed for ``good'' and ``ok'' fields (defined
below).  The calibrations are better for the 1.3m, with tßhe
distribution peaking around 0.018 mag.

\section{Results}

\subsection{Survey Coverage and Depth}

Individual exposures were assigned one of three quality ratings.
``Good'' fields were taken under photometric conditions, in reasonable
seeing (less than 2.5 arcsec for the Bok survey, less than 3 arcsec
for the 1.3m survey), have acceptable pointing errors,
have astrometric calibration errors of less than 60 mas in both right
ascension and declination, and have a minimum of 0.09 square degrees
of SDSS coverage.  ``Ok'' fields have the same requirements on the
astrometric calibration errors and SDSS coverage, but can have up to
half a magnitude of extinction due to clouds, seeing up to 4 arcsecs,
and have no restrictions on pointing errors.  The remaining
observations are declared ``bad'', and are not included in the
catalog.  For the Bok, 78.3\% of the observations were ``good'',
14.7\% were ``ok'', and 7.0\% were ``bad''.  For the 1.3m, 67.7\% of
the observations were ``good'', 15.4\% were ``ok'', and 16.9\% were
``bad''.

\begin{figure}
\plotone{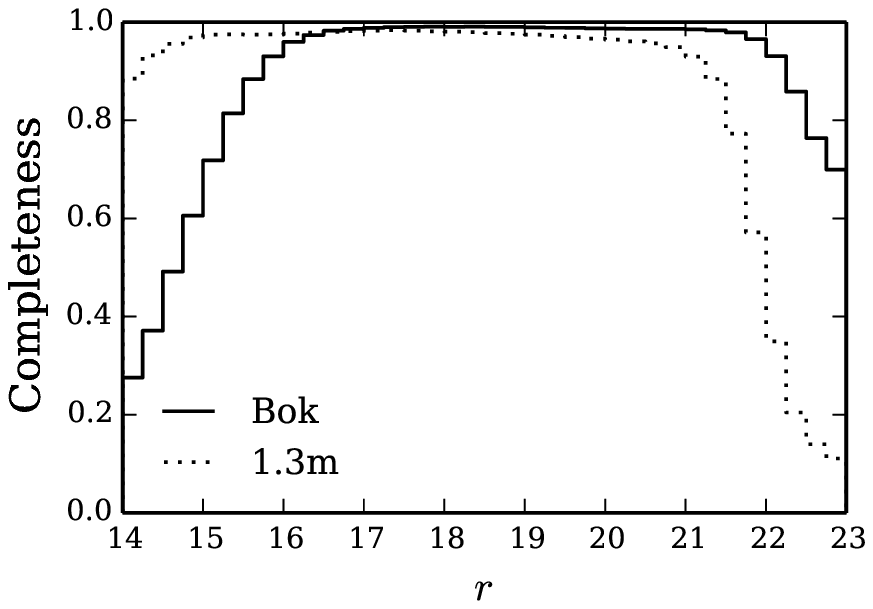}
\caption{Completeness in the ``good'' fields for the Bok (solid
  histogram) and 1.3m (dotted histogram) surveys.}
\label{fig-completeness}
\end{figure}

The $r$ magnitude at which a given field reaches 90\% completeness is
well approximated for the Bok by the formula
\begin{eqnarray}
r_{90} & = & 22.33 - 0.5 (\rm{seeing} - 1.5) + 0.5 (\rm{sky} - 21.03)\\
\nonumber
& & + (\rm{zeropt} - 6.40),
\end{eqnarray}
and for the 1.3m by the formula
\begin{eqnarray}
r_{90} & = & 21.61 - 0.5 (\rm{seeing} - 2.0) + 0.5 (\rm{sky} - 20.48)\\
\nonumber
& & + (\rm{zeropt} - 4.57),
\end{eqnarray}
where ``seeing'' is the seeing in arcsecs, ``sky'' is the sky brightness
in mag~arcsec$^{-2}$, and ``zeropt'' is the photometric zero point, which serves
as a measure of the extinction due to clouds.  The Bok benefits from both
better seeing than the 1.3m (typical seeing of 1.5 arcsec versus 2.0 arcsec)
and darker skies (typical sky brightness of 21.0~mag~arcsec$^{-2}$ versus
20.5~mag~arcsec$^{-2}$), leading to the Bok reaching about a magnitude fainter
than the 1.3m.
Figure~\ref{fig-completeness} displays the mean completeness versus
magnitude for the ``good'' fields in both surveys.  The Bok survey
is 90\% complete to 22.3, and 95\% complete to 22.0.  The 1.3m survey
is 90\% complete to 21.3, and 95\% complete to 20.9.  The incompleteness
at the bright end reflects the onset of saturation at $r = 16$ for the Bok
and $r = 15$ for the 1.3m.

\begin{figure}
\plotone{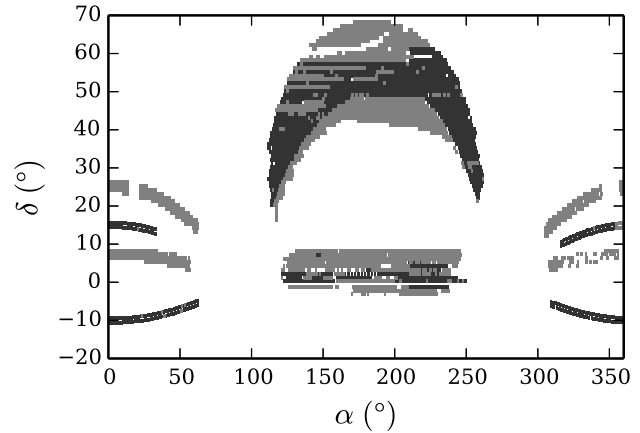}
\caption{Sky coverage of the survey, including both ``good'' and ``ok''
  observations.  Dark and light regions indicate Bok and 1.3m coverage,
  respectively.}
\label{fig-coverage}
\end{figure}

Some target fields have more than one observation.
In those cases, the observation with the deepest estimated completeness is
included in the survey.  The set of ``good'' fields, of fairly uniform
depth, comprises the primary product of this survey, with the set of
``ok'' fields comprising a secondary product of varying and lesser depth.
The ``good'' fields from the Bok survey cover 1097.9 square degrees of sky,
while the ``ok'' fields add another 145.6 square degrees.
The ``good'' fields from the 1.3m survey cover 1521.4 square degrees of sky,
while the ``ok'' fields add another 342.6 square degrees.  The sky coverage
is displayed in Figure~\ref{fig-coverage}.

\subsection{Proper Motions}

QSOs are morphologically stellar, yet distant enough to have
undetectable proper motions to the accuracy of this survey.  The
distribution of measured proper motions for a sample of QSOs thus
provides a simple test of the accuracy of the measured proper motions.
All analysis in this section uses the SDSS Data Release 7 catalog of
spectroscopically confirmed QSOs \citep{schneider2010}.  The QSO
sample is limited to those with clean detections in both SDSS and this
survey, by requiring that (1) the SDSS detection meets the criteria
for a ``clean'' point source as recommended on the SDSS DR7
website;
(2) it has a clean detection in this survey as indicated by the
SExtractor flags; (3) its observation in this survey has a quality of
``good''; and (4) it has an epoch difference of at least 4 years.
This yields a sample of 10855 QSOs in the
Bok survey footprint, and 12621 in the 1.3m survey footprint.

\begin{figure}
\plotone{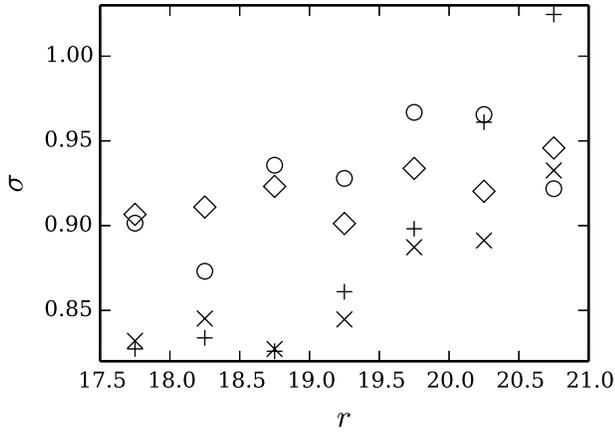}
\caption{Standard deviation of the distribution of QSO proper motions
  divided by proper motion errors, in half-magnitude bins.  Circles
  and diamonds are the standard deviations in right ascension and
  declination, respectively, for the Bok survey.  Pluses and
  crosses are the standard deviations in right ascension and
  declination, respectively, for the 1.3m survey.  Perfectly estimated
  errors would yield a standard deviation of one.}
\label{fig-qso-std}
\end{figure}

The accuracy of the proper motion error estimates is first examined by
binning the QSOs in half magnitude bins in the range $17.5 < r <21$,
and within each bin examining the distribution of the proper motions,
divided by the errors in the proper motions, separately in right
ascension and declination.  If the error estimates are accurate the
distribution in each bin should be a Gaussian with a standard
deviation of one.  The distribution in each bin is well represented by
a Gaussian.  Figure~\ref{fig-qso-std} plots the standard deviation of
the fitted Gaussian in each bin against $r$ magnitude.  The errors are
overestimated by 10 --- 15\% at bright magnitudes, becoming more
accurate towards fainter magnitudes.  The calibration errors are thus
likely somewhat overestimated, as they dominate the errors at bright
magnitudes (the centering errors become comparable to the
calibration errors around $r \sim 21$).

\begin{figure}
\plotone{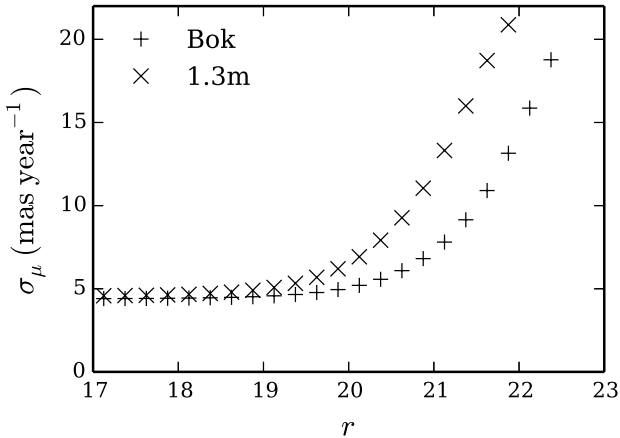}
\caption{Mean proper motion error (averaged over the right ascension and
  declination components), scaled to an epoch difference of 6 years, as a
  function of $r$ magnitude, for a sample of clean stars with a field
  quality of ``good''.  Pluses and crosses are for the Bok and 1.3m
  surveys, respectively.}
\label{fig-errors-by-mag}
\end{figure}

The dependence of proper motion errors on magnitude is shown in
Figure~\ref{fig-errors-by-mag}.  The mean proper motion errors
(averaged over the right ascension and declination components) for a
sample of clean stars in fields of quality ``good'', scaled to a
typical epoch difference of 6 years, are plotted in quarter magnitude
bins.  The proper motion errors scale inversely with epoch difference.
The greater depth of the Bok survey is evident, obtaining comparable
proper motion errors roughly 0.75 mag fainter than the 1.3m survey.
At their 95\% completeness limits ($r = 22.0$ for the Bok survey, and
$r = 20.9$ for 1.3m survey), both surveys have mean proper motion
errors of about 15~\masyear\ (for an epoch difference of six years).

\begin{figure*}
\plotone{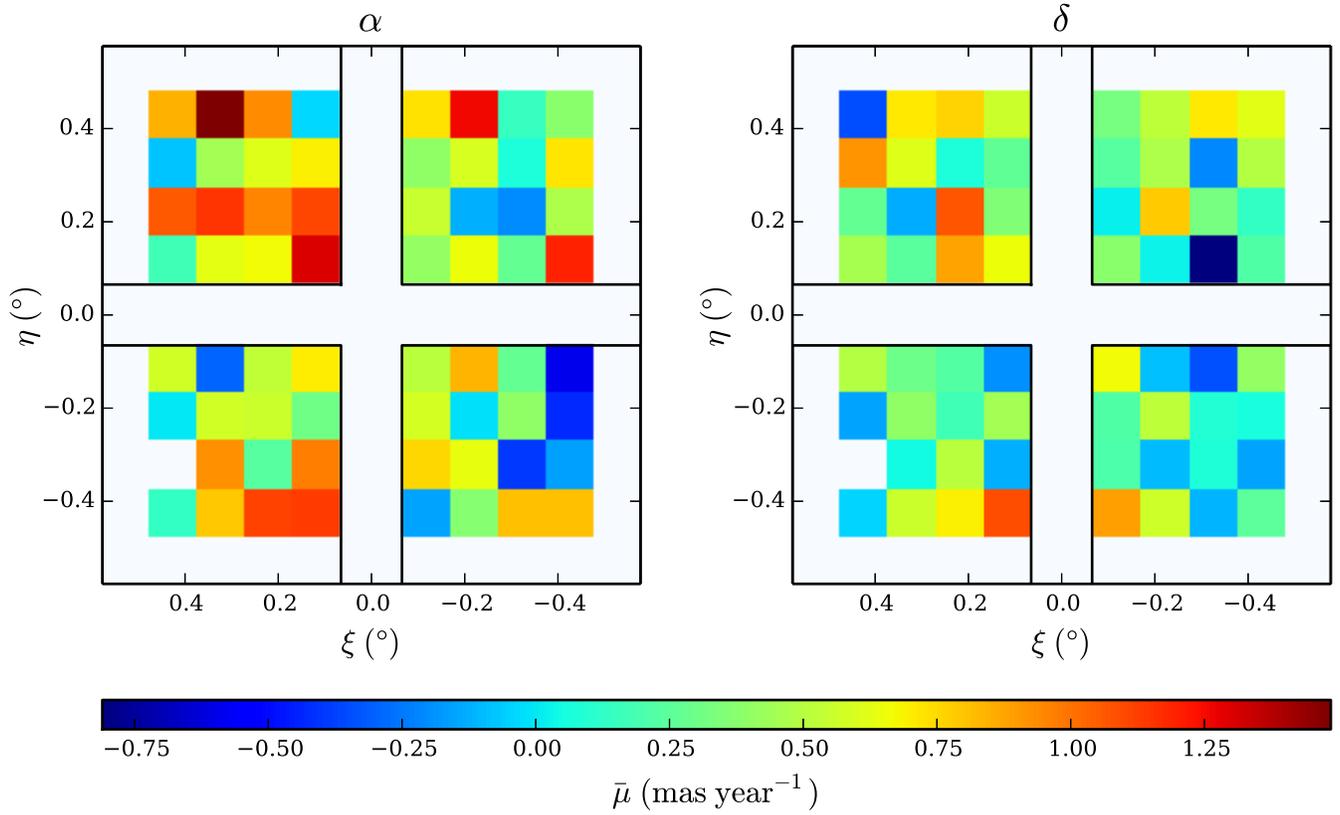}
\caption{Mean proper motion of QSOs binned across the Bok focal plane.
  Proper motions in right ascension are shown in the left figure,
  and in declination in the right figure.  The white bins along the
  edges of the focal plane have too few QSOs for a proper measure, due
  to overlap between fields.}
\label{fig-fp-bok-mean}
\end{figure*}

\begin{figure*}
\plotone{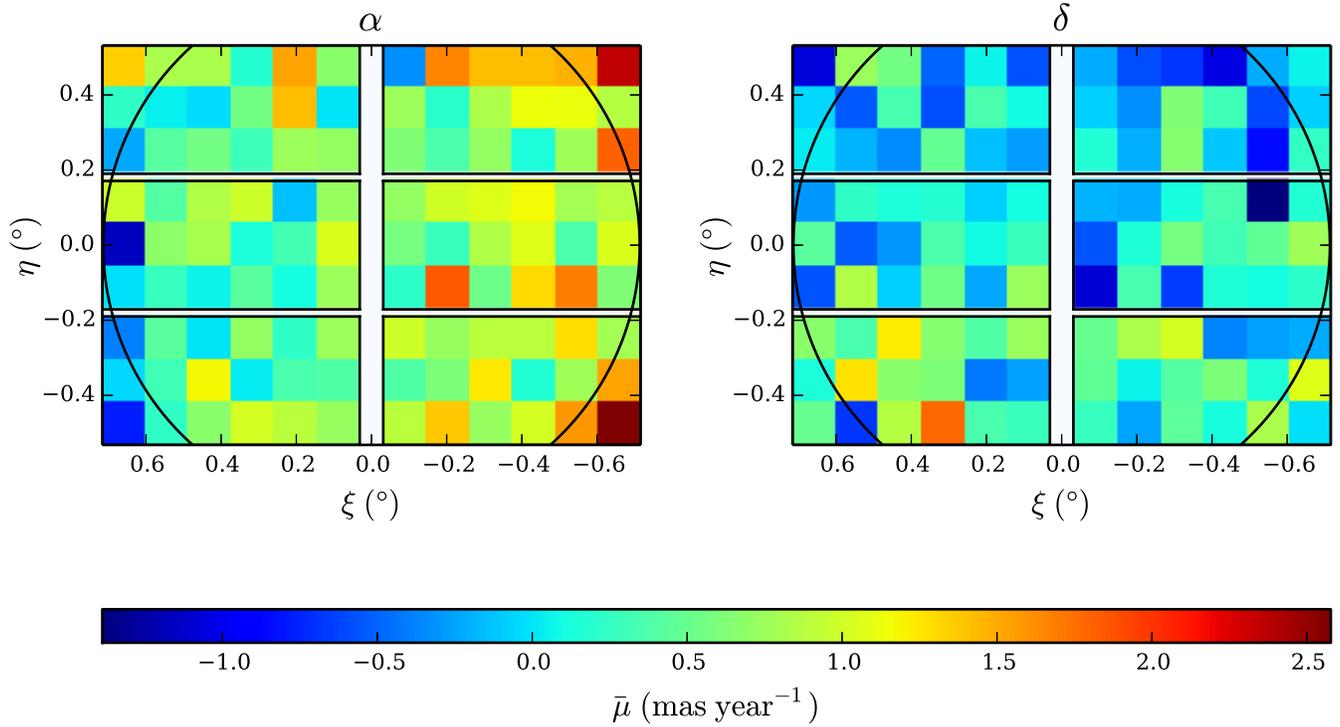}
\caption{Mean proper motion of QSOs binned across the 1.3m focal plane.
  Proper motions in right ascension are shown in the left figure,
  and in declination in the right figure.  A circle with radius 0.7
  degrees is shown for reference.}
\label{fig-fp-13m-mean}
\end{figure*}

\begin{figure*}
\plotone{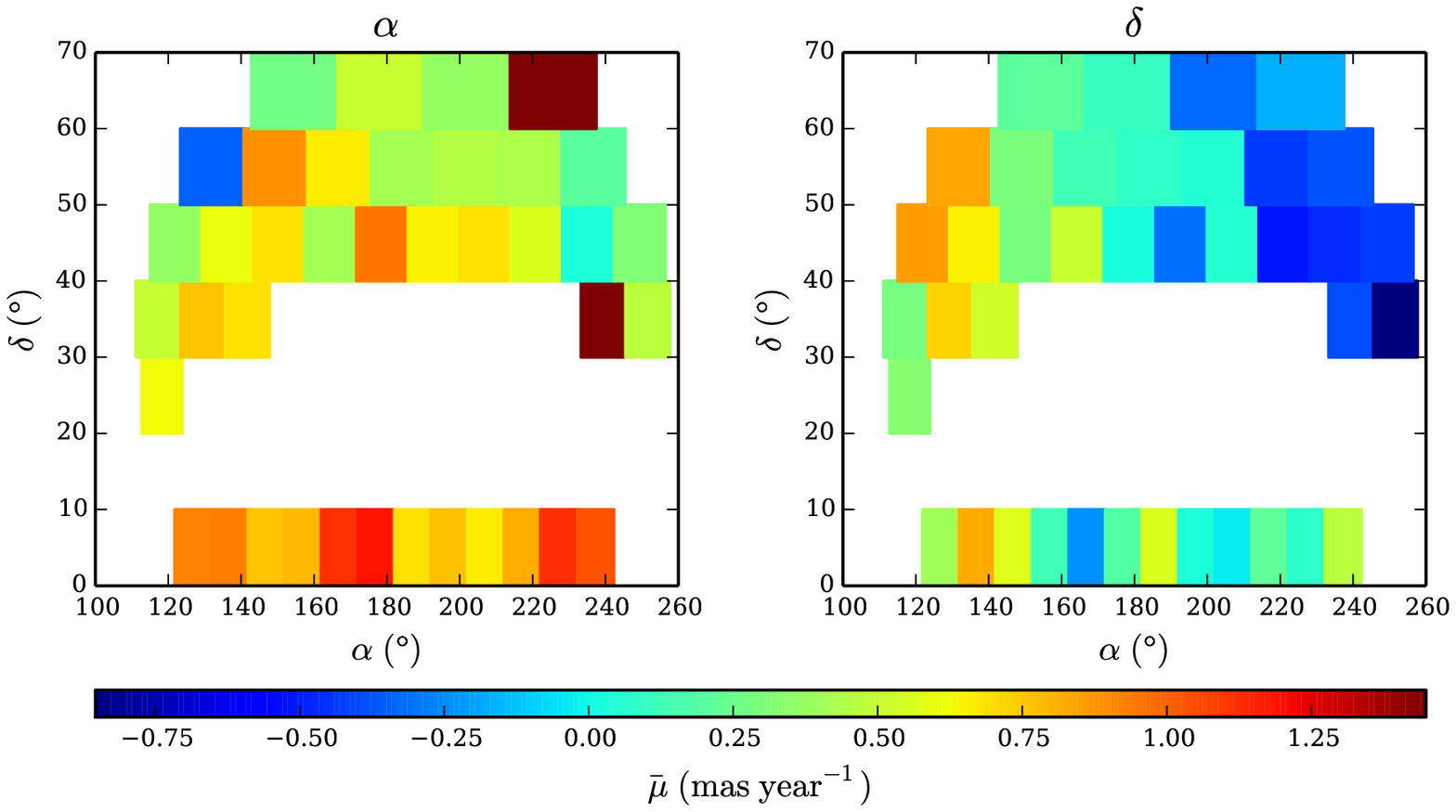}
\caption{Mean proper motion of QSOs in 100 square degree bins in the north
  galactic cap region of the survey.  Proper motions in right ascension
  are shown in the left figure, and in declination in the right figure.}
\label{fig-qsos-sky}
\end{figure*}

The mean proper motion for all QSOs in the Bok survey is
0.5~\masyear\ in right ascension and 0.3~\masyear\ in
declination, and for the 1.3m survey 0.7~\masyear\ in right
ascension and 0.1~\masyear\ in declination.  The expected error
in the mean is about 0.05~\masyear, thus the mean motions
represent real systematic errors.  There is no magnitude dependence
for the mean motions.  0.6\% of the Bok QSOs, and 0.4\% of the 1.3m
QSOs, have proper motions greater than three times their proper motion
errors, compared with the 0.3\% expected for a normal distribution.
Figures~\ref{fig-fp-bok-mean} and \ref{fig-fp-13m-mean} show the
distribution of mean proper motions across the focal plane (the
expected error in the mean for each bin is around
0.5~\masyear).  There are no evident systematics across either
focal plane.  Figure~\ref{fig-qsos-sky} displays the distribution of
mean proper motions across the northern galactic cap region of the
survey, binned in 100 square degree segments of sky.  Each bin
contains more than 300 stars, yielding an expected error in the mean
motions of less than 0.3~\masyear.  There are some large scale
systematic mean motions, particularly in declination, of up to
1~\masyear.  DCR effects are one possible source of such
systematics, as QSOs have very different spectral energy
distributions from stars, and the fields at low and high right
ascension within the north galactic cap region tend to be observed
further from the meridian and at higher zenith distance.  The
SDSS+USNO-B survey shows similar systematics \citep{bond2010},
and these will propagate to our survey since the SDSS+USNO-B proper
motions are used to propagate the SDSS positions to the survey epoch.

The most complete, deep, wide-area proper motion catalog available for
comparison is the LSPM catalog (\citealt{lspm}; we don't use the
SDSS+USNO-B catalog for this purpose, since it provides our
astrometric calibrators, nor the USNO-B catalog, since it has a large
contamination rate at large proper motions).  The catalog is based on
the Space Telescope Science Institute scans of the POSS-I and POSS-II
Schmidt plates.  It improves on other high-proper-motion catalogs
based on the POSS plates in using an image subtraction technique,
which allows them to achieve an estimated completeness of 99\% at high
Galactic latitudes for $V < 19$ and proper motions larger than
0.15\arcsec~year$^{-1}$ (the current published catalog is for stars
north of the celestial equator).  We can thus measure our completeness
by comparison with LSPM stars in the magnitude range $16 < V < 19$,
where the bright limit avoids saturated stars in our survey, and the
faint limit corresponds to the faint limit of the LSPM survey.  There
are 5410 LSPM stars in that magnitude range in our survey footprint,
of which all but 114 have a matching catalog entry in our survey, for
a completeness rate of 97.9\%.  The completeness is a function of
total proper motion, falling from 99\% for a total proper motion of
150~\masyear\ to 94\% for a total proper motion of 700~\masyear.  Three of
the unmatched LSPM stars are false detections in LSPM, which becomes
apparent with the two additional CCD epochs.  Another 99 of the unmatched LSPM
stars are close pairs of detections in the SDSS survey but single
(unresolved) detections in our survey, in which the LSPM star is
matched to one detection in the SDSS pair, and our detection is
matched to the other SDSS detection.  Of these 99 cases, 51 are
clearly pairs of blended stars which are resolved as two stars in the
SDSS catalog, but unresolved in our survey, while the remaining appear
by eye to be multiple detections of the same single star in the SDSS
image, though it's certainly possible that they are indeed blended
pairs of stars resolved by the SDSS object detection algorithm but too
close to be obviously two stars by eye inspection.  One LSPM star falls
in an apparent hole in the SDSS data which is not listed in the DR7
hole list.  The remaining 11 unmatched LSPM stars are located near
very bright stars, and undetected in our survey either because they
were lost in the scattered light of the bright star, or were blended
with a diffraction spike from the star.

For the LSPM stars that have matching entries in our catalog, 188 have
proper motions in our catalog which disagree with the LSPM proper
motions by more than three times the expected errors.  Of these, 24 are
false detections in the LSPM catalog, which again is apparent with the
additional CCD epochs.  For an additional 77, our
proper motions are clearly better than the LSPM proper motions, due
primarily to better resolution of blended stars or deeper exposures of
faint stars on the CCD images versus the Schmidt plates.  Thus, only 87
of our proper motions are suspect, for a contamination rate of 1.6\%.
This rate may be an overestimate, as for 20 of the 87, it is not clear
by eye inspections of the images that either our or the LSPM proper
motions are wrong.  The remaining 67 objects clearly have errant
proper motions in our catalog.  Of these, three have problems with the
SDSS detection, while the remaining 64 are errant detections
in our survey, due primarily to blends with other stars or diffraction
spikes, mismatches, or poor centers.

\begin{figure}
\plotone{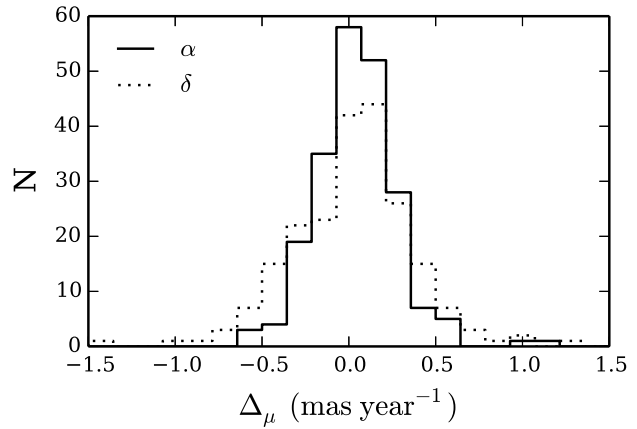}
\caption{Distribution of mean differences between Bok and 1.3m proper
  motions in individual duplicate fields.  Solid and dotted histograms are
  the distributions for proper motions in right ascension and
  declination, respectively.}
\label{fig-bok-13m-diff}
\end{figure}

There are 214 pairs of observations where the same target field was
observed with both the Bok and 1.3m telescopes, and each observation
in the pair was of ``good'' quality.  These cover 130 square degrees
of the survey, providing repeat observations of 536,630 SDSS stars
in the magnitude range $16 < r
< 21.5$, which can be further used to examine the survey completeness
and contamination.  A total of 98.4\% of the stars were detected in
the Bok observations, 97.5\% were detected in the 1.3m observations,
and 96.8\% were detected in both the Bok and 1.3m surveys.  For those
stars detected in both surveys, 98.5\% have proper motions which agree
within three times the expected errors.  Assuming these proper motions
are thus correct (in some cases we can certainly have matching yet
incorrect proper motions, such as the same mismatch in both surveys,
however based on the comparisons with LSPM discussed above these
should be rare), this yields a contamination rate of 1.5\%, in good
agreement with the contamination rate based on the LSPM analysis.
Figure~\ref{fig-bok-13m-diff} displays histograms of the mean
differences between the Bok and 1.3m proper motions in individual
fields.  Systematic differences are consistently less than
0.5~\masyear.

For a survey as large as this, even a 1.5\% contamination rate yields
a large number of objects with incorrect proper motions, and the
contamination rate is expected to increase with increasing proper
motion.  The primary science driver for the catalog is to generate a
large statistically well-defined sample of faint proper motions for
Galactic structure studies.  We are more interested in pushing the
catalog to small proper motions, so as to increase the sample size and
depth, than we are in generating a complete and contamination free
sample of high proper motion stars.  The latter is beyond the scope of
the paper.  However we can use both the LSPM matches and duplicate
detections to help isolate sets of cuts to the catalog so as to remove
some fraction of contaminants while maintaining a high completion
rate.  Table~\ref{table-cuts} presents such a set of cuts.  For each
cut, we list the fraction of our stars that we consider as good or
bad detections, based on matches with the LSPM catalog, which meet the
cut.  Similar results are listed based on our analysis of duplicate
detections between the Bok and 1.3m surveys.  An ideal cut would
retain most of the good objects, while rejecting a large fraction of
the bad objects.  While all of the proposed cuts retain at least 97\%
of the good objects, none reject more than half of the bad objects.
The PSF fitting is adversely affected by the fixed pattern noise on
CCD 2 in the Bok survey.  In cases where statistics vary significantly
between the Bok CCD 2 data and the rest of the survey, the statistics
for the Bok CCD 2 data are given in the footnotes, and the data for
the rest of the survey are given in the table.  It is also the case
that the worst of the FOV-dependent systematic errors in the proper
motions for the Bok survey occur on CCD 2.  Those applications which
require the smallest systematics may wish to exclude the data from CCD
2 for the Bok survey.

\begin{deluxetable}{lrrrrrrrr}
\tablecaption{Catalog Cuts\label{table-cuts}} \tablehead{ \colhead{} &
  \multicolumn{2}{c}{LSPM} & \multicolumn{2}{c}{Duplicates}
  \\ \colhead{} & \colhead{Good} & \colhead{Bad} & \colhead{Good} &
  \colhead{Bad} \\ \colhead{Cut} & \colhead{(\%)} & \colhead{(\%)} &
  \colhead{(\%)} & \colhead{(\%)} } \startdata Good SDSS
detection\tablenotemark{a} & 100.0 & 91.9 & 100.0 & 100.0
\\ SExtractor detection\tablenotemark{b} & 98.4 & 66.6 & 98.4 & 76.8
\\ 1-to-1 match\tablenotemark{c} & 98.3 & 87.7 & 100.0 & 83.8
\\ $\rm{nIter} < 10$\tablenotemark{d} & 98.9\tablenotemark{e} &
59.5 & 99.1 & 53.1 \\
$\rm{chi} < 8$\tablenotemark{f} & 97.6\tablenotemark{g} & 57.8 &
99.2\tablenotemark{h} & 71.3 \\
$|\Delta_r| <0.5$\tablenotemark{i} & 99.4\tablenotemark{j} & 61.6 & 99.4 & 59.4
\enddata
\tablenotetext{a}{A clean SDSS detection, requiring the SDSS flag
combination BINNED1 \& !(BRIGHT $|$ NOPROFILE $|$ DEBLEND\_NOPEAK).}
!
\tablenotetext{b}{Detected using SExtractor.}
\tablenotetext{c}{Exactly one object in this survey matched exactly
  one SDSS object.}
\tablenotetext{d}{The number of iterations when
  fitting this object to the PSF in DAOPHOT.  10 was the maximum
  number of iterations allowed, thus objects with $\rm{nIter} = 10$
  were not well fit by the PSF.} 
\tablenotetext{e}{Excludes Bok CCD 2; 96.1\% for Bok CCD 2.} 
\tablenotetext{f}{DAOPHOT chi parameter.}
\tablenotetext{g}{Excludes Bok CCD 2; 62.8\% for Bok CCD 2.}
\tablenotetext{h}{Excludes Bok CCD 2; 89.2\% for Bok CCD 2.}
\tablenotetext{i}{Absolute difference between the SDSS $r$ magnitude
  and our magnitude.}  \tablenotetext{j}{Excludes Bok CCD 2; 96.5\%
  for Bok CCD 2.}
\end{deluxetable}

\subsection{Photometry}

\begin{figure}
\plotone{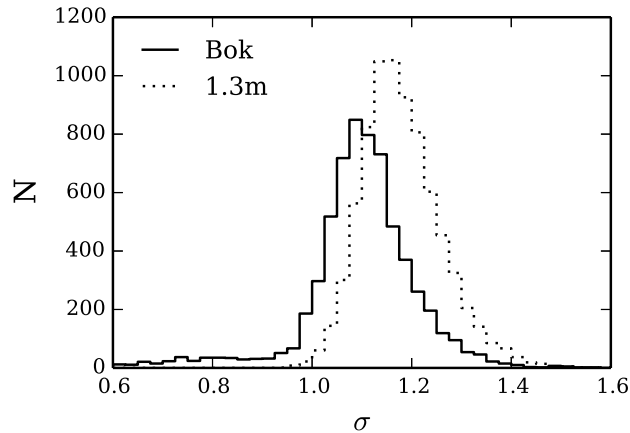}
\caption{Distribution of the standard deviation in each field of the
  difference between SDSS and our magnitudes, normalized by the
  expected errors in the differences, for ``good'' fields only.  Solid
  and dotted histograms are for the Bok and 1.3m surveys, respectively.
  Perfectly estimated errors would yield a standard deviation of one.}
\label{fig-pcalib-field-rms}
\end{figure}

\begin{figure}
\plotone{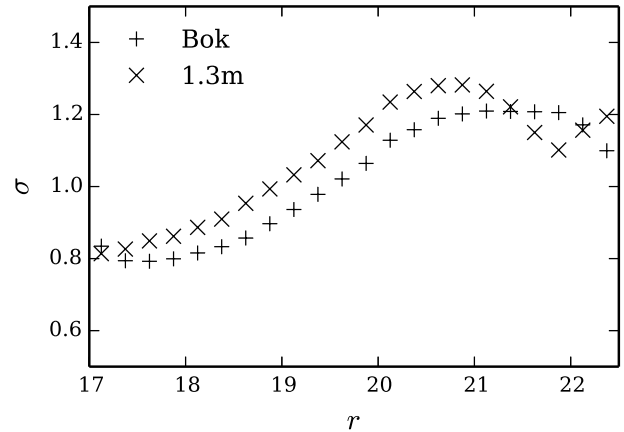}
\caption{Mean standard deviation of the difference between SDSS and
  our magnitudes, normalized by the expected errors in the
  differences, for ``good'' fields only, versus $r$ magnitude.  Pluses and
  crosses are the standard deviations for the Bok and 1.3m
  surveys, respectively.  Perfectly estimated errors would yield a
  standard deviation of one.}
\label{fig-pcalib-mag-rms}
\end{figure}

We examine the accuracy of the photometry error estimates by examining
the distribution of magnitude differences between the SDSS magnitudes
and our magnitudes, divided by the estimated errors in the
differences.  If our (and SDSS's) error estimates are accurate, the
distributions should be Gaussian with standard deviations of one.
Figure~\ref{fig-pcalib-field-rms} displays histograms of the
standard deviations of normalized errors for each image of quality
``good'', for stars in the magnitude range $17 < r < 21.5$.
Typically, the Bok errors are underestimated by about 10\%, and the
1.3m errors by about 15\%.  The Bok distribution shows a tail of images
whose errors are overestimated by 10 -- 40\%; these are all from CCD 2,
and are an effect of the fixed pattern noise on that CCD.
Figure~\ref{fig-pcalib-mag-rms} displays
the mean standard deviation in half magnitude bins, separately for each
survey.  Both surveys overestimate the errors at the bright end by about
20\%, indicating likely overestimated calibration errors, while underestimating
at the faint end by about 20\%.

\begin{figure}
\plotone{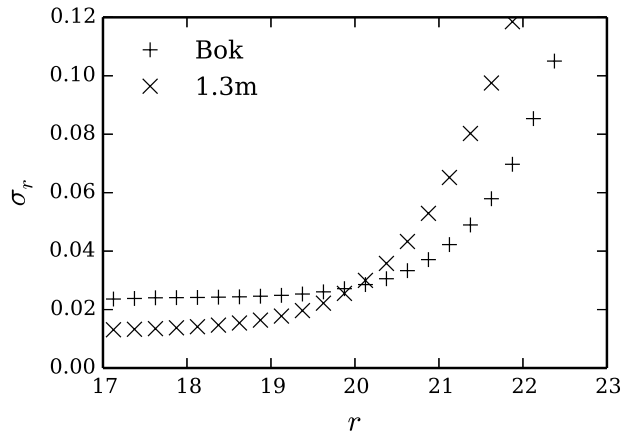}
\caption{Mean photometric error as a
  function of $r$ magnitude, for a sample of clean stars with a field
  quality of ``good''.  Pluses and crosses are for the Bok and 1.3m
  surveys, respectively.}
\label{fig-perrors-by-mag}
\end{figure}

The dependence of photometric errors on magnitude is shown in
Figure~\ref{fig-perrors-by-mag}.  The mean photometric errors for a
sample of clean stars in fields of quality ``good'' are plotted in
quarter magnitude bins.  At their 95\% completeness limits ($r = 22.0$
for the Bok survey, and $r = 20.9$ for 1.3m survey), both surveys have
mean photometric errors of about 0.07 magnitudes.

\section{The Catalog}

The observations reported here add proper motions and a second epoch
of $r$ band photometry to a portion of the SDSS DR7 catalog.  Any
science conducted with the catalog will rely on SDSS DR7 for its
precise 5-band photometry, morphological classification, extinction
estimates, etc.  Rather than duplicate the SDSS DR7 data, we make our
catalog available within the SDSS Catalog Archive Server Jobs System
(CasJobs)\footnote{http://skyserver.sdss.org/casjobs/}.  The data are
contained in four tables in the special user ``deepPM'''s MyDB that
are published to the ``public'' group so that they are available to all
users of the CasJobs system. These tables have links to the matching
SDSS objects in the DR7 context (see sample query below for the syntax
to reference these tables and link them to the DR7 data).

The first table, ``Observation'', has an entry for each observation in
the survey (excluding ``bad'' observations).  The schema is given in
Table~\ref{table-observation}.  The area of sky covered by each image
is the intersection of the CCD footprint and SDSS areal coverage,
excluding bad regions of the CCD.  We have used
MANGLE\footnote{http://space.mit.edu/$\sim$molly/mangle/}
\citep{mangle1, mangle2} to define the sky coverage for each image.
Further, we have adopted the bright star masks of \citet[][reproducing
  the algorithm for some of the southern SDSS stripes not included in
  that paper]{blanton2005} to optionally exclude regions affected by
bright stars.  In the ``Observation'' table, the attributes
``fullAreaX'' give the full areal coverage for CCD X, while ``areaX''
gives the areal coverage exluding the bright star masks.  The
spherical polygons comprising the MANGLE descriptions of the sky
coverage for each image are given in the table ``Cap'', whose schema
is given in Table~\ref{table-cap}.  The nomenclature follows that of
MANGLE's ``polygon'' file format.  Each image (identified by the
combination of columns ``night'', ``obsID'', and ``ccd'') is covered
by the union of one or more polygons, and each polygon is defined by
the intersection of spherical caps.  Two separate sets of polygons are
given for each image, one which excludes regions affected by bright
stars, and one which does not.

Each SDSS DR7 star or galaxy that falls within our survey area, and
within a given $r$ magnitude range, is listed in the tables ``PMStar''
or ``PMGalaxy'', respectively.  For the Bok survey, we include SDSS
objects in the magnitude range $16 \leq r \leq 23$, while for the 1.3m
survey we include objects in the magnitude range $15 \leq r \leq 22$
(using PSF magnitudes for stars and model magnitudes for galaxies).
The bright limits are determined by the onset of saturation for stars,
and the faint limits are set to one magnitude below the approximate
95\% completeness limits.  There is a table entry for each SDSS
object, whether or not it was detected in our survey.  Unmatched SDSS
objects will have the column ``match'' set to 0.  Objects in our
survey are matched to SDSS objects by searching in annuli using
progressively larger radii.  If more than one match is found within an
annulus, the nearest match is used.  While galaxies will not have
detectable proper motions in our survey, we list proper motions for
galaxies because the SDSS star/galaxy classification is not perfect,
particularly at the faint end of our survey, and some users may choose
to use their own morphological classifications.  The measured proper
motions of galaxies also provide additional quality analysis, in the
same way the measured proper motions of QSOs do.  The schema for the
``PMStar'' and ``PMGalaxy'' tables is given in
Table~\ref{table-object}.  In addition to links to the matching SDSS
objects, and to the observations in our survey in which the SDSS
objects should have been detected, the tables list the proper motions
in right ascension and declination, their uncertainties, the $r$ magnitudes
in our survey (derived from the PSF fits, even for galaxies) and their
uncertainties, various parameters characterizing the PSF fits as produced by
DAOPHOT, and various image parameters produced by SExtractor.  Some
matched objects lack either DAOPHOT or SExtractor (but not both)
measurements, in which case those attributes will be set to 0.  Only
objects successfully measured by DAOPHOT have measured proper motions,
since the proper motions are based on the DAOPHOT centers.  The
``PMStar'' table contains 21,157,643 entries, of which 18,982,227 have
measured proper motions in our survey, and the ``PMGalaxy'' table
contains 33,782,162 entries, of which 25,506,247 have measured proper
motions in our survey.

The tables are public tables in the SDSS CasJobs.  Refer to the
CasJobs documentation for instructions on accessing public tables.  As
an an example of accessing the data, executing the following batch SQL
query within the DR7 context in CasJobs will return SDSS positions and
$ugriz$ photometry, along with our proper motions, for all stars in
the Bok survey that: (1) have a ``good'' observation in our survey,
(2) have a one-to-one match between SDSS and our survey, (3) are in
the magnitude range $20 < r < 22$, and (4) have a proper motion
greater than 100~\masyear:
\begin{verbatim}
  SELECT s.ra, s.dec, s.psfMag_u, s.psfMag_g,
         s.psfMag_r, s.psfMag_i, s.psfMag_z,
         p.pmRa, p.pmDec
  FROM public.deepPM.PMStar p
  JOIN public.deepPM.Observation o ON
       p.night = o.night AND p.obsID = o.obsID
  JOIN Star s ON p.objID = s.objID
  WHERE o.survey = 0 AND
        o.quality = 1 AND
        p.match = 11 AND
        s.psfMag_r BETWEEN 20 and 22 AND
        p.pmRa * p.pmRa + p.pmDec * p.pmDec > 
        100 * 100
\end{verbatim}

\begin{deluxetable*}{llll}
\tablecaption{Observation Schema\label{table-observation}}
\tablehead{
  \colhead{Name} & \colhead{Type} & \colhead{Units} & \colhead{Description}
}
\startdata
night\tablenotemark{a} & int32 & & MJD number of the night the observation was obtained. \\
obsID\tablenotemark{a} & int16 & & Observation number, unique within a given night. \\
survey & int8 & & Which survey is this observation part of: 0=Bok, 1=1.3m. \\
mjd & float64 & & MJD at start of the observation. \\
expTime & float32 & seconds & Exposure time. \\
ra & float64 & degrees & Right ascension of target field center. \\
dec & float64 & degrees & Declination of target field center. \\
fwhm & float32 & arcsecs & Average FWHM over all CCDs. \\
sky & float32 & mags arcsec$^{-2}$ & Average sky brightness over all CCDs. \\
zero & float32 & mags & Average photometric zeropoint over all CCDs. \\
raRms & float32 & arcsecs & Average astrometric calibration rms residual in right ascension over all CCDs. \\
decRms & float32 & arcsecs & Average astrometric calibration rms residual in declination over all CCDs. \\
quality & int8 & & Data quality: 1=good, 0=ok. \\
ccd1 & int8 & & Is data for CCD 1 included in the survey: 1=yes, 0=no. \\
fullArea1 & float32 & degrees$^{2}$ & Area of sky covered by CCD 1, including regions affected by bright stars. \\
area1 & float32 & degrees$^{2}$ & Area of sky covered by CCD 1, excluding regions affected by bright stars. \\
minDT1 & float32 & years & Minimum epoch difference for all objects on CCD 1. \\
fwhm1 & float32 & arcsecs & FWHM on CCD 1. \\
sky1 & float32 & mags arcsec$^{-2}$ & Sky brightness on CCD 1. \\
zero1 & float32 & mags & Photometric zeropoint on CCD 1. \\
raRms1 & float32 & arcsecs & Astrometric calibration rms residual in right ascension on CCD 1. \\
decRms1 & float32 & arcsecs & Astrometric calibration rms residual in declination on CCD 1. \\
psfRms1 & float32 & mags & Photometric calibration rms residual on CCD 1. \\
ccd2 & int8 & & Is data for CCD 2 included in the survey: 1=yes, 0=no. \\
fullArea2 & float32 & degrees$^{2}$ & Area of sky covered by CCD 2, including regions affected by bright stars. \\
area2 & float32 & degrees$^{2}$ & Area of sky covered by CCD 2, excluding regions affected by bright stars. \\
minDT2 & float32 & years & Minimum epoch difference for all objects on CCD 2. \\
fwhm2 & float32 & arcsecs & FWHM on CCD 2. \\
sky2 & float32 & mags arcsec$^{-2}$ & Sky brightness on CCD 2. \\
zero2 & float32 & mags & Photometric zeropoint on CCD 2. \\
raRms2 & float32 & arcsecs & Astrometric calibration rms residual in right ascension on CCD 2. \\
decRms2 & float32 & arcsecs & Astrometric calibration rms residual in declination on CCD 2. \\
psfRms2 & float32 & mags & Photometric calibration rms residual on CCD 2. \\
...\tablenotemark{b}
\enddata
\tablenotetext{a}{The combination of columns ``night'' and ``obsID'' uniquely identify an observation, and together comprise the primary key for the table.}
\tablenotetext{b}{The attributes ``ccdX'' -- ``psfRmsX'' are repeated for CCDs (X) 3 -- 6.  For the Bok survey,  the attributes for CCD 5 and 6 will all be set to 0, since the camera only has 4 CCDs.}
\end{deluxetable*}

\begin{deluxetable*}{llll}
\tablecaption{Cap Schema\label{table-cap}}
\tablehead{
  \colhead{Name} & \colhead{Type} & \colhead{Units} & \colhead{Description}
}
\startdata
night & int32 & & MJD number of the night the observation was obtained.\\
obsID & int16 & & Observation number.\\
ccd & int8 & & CCD covered by this cap.\\
brightStar & int8 & & 1=excludes regions affected by bright star masks, 0=doesn't.\\
polygon & int8 & & Identifier of polygon to which this cap belongs.\\
x & float64 & & X component of the unit vector defining the north polar axis of the cap.\\
y & float64 & & Y component of the unit vector defining the north polar axis of the cap.\\
z & float64 & & Z component of the unit vector defining the north polar axis of the cap.\\
cm & float64 & & 1 - cos($\theta$), where $\theta$ is the polar angle of the cap.\\
& & & Positive/negative cm designates the region north/south of the polar angle. 
\enddata
\end{deluxetable*}

\begin{deluxetable*}{llll}
\tablecaption{PMStar/PMGalaxy Schema\label{table-object}}
\tablehead{
  \colhead{Name} & \colhead{Type} & \colhead{Units} & \colhead{Description}
}
\startdata
objID & int64 & & Unique SDSS identifier for the SDSS object. \\
night & int32 & & MJD number of the night the observation was obtained. \\
obsID & int16 & & Observation number. \\
ccd & int8 & & CCD on which the object was or should have been detected. \\
brightStar & int8 & & 1 if objects falls in a masked region around a bright star, else 0. \\
match & int16 & & Multiplicity of matches between our survey and SDSS.  Ones
digit indicates\\
& & & number of matching objects in SDSS, and 10s digit indicates number of \\
& & &  matching objects in this survey.  Thus, 11 indicates a one-to-one match.\\
& & & If 0, there was no match for this SDSS object in our survey, and all remaining\\
& & & columns for this object will be set to 0. \\
pmRa & float32 & mas~year$^{-1}$ &Proper motion in right ascension, along the great circle (i.e., $\dot{\alpha}\cos(\delta)$). \\
pmRaErr & float32 & mas~year$^{-1}$ & Error in proper motion in right ascension. \\
pmDec & float32 & mas~year$^{-1}$ & Proper motion in declination. \\
pmDecErr & float32 & mas~year$^{-1}$ & Error in proper motion in declination. \\
r & float32 & mags & $r$ magnitude in our survey. \\
rErr & float32 & mags & Error in $r$ magnitude. \\
x & float32 & pixels & X-axis coordinate of object center. \\
y & float32 & pixels & Y-axis coordinate of object center. \\
nIter & int8 & & Number of iterations when DAOPHOT fit the PSF.  The maximum is 10 \\
& & & iterations, thus a value of 10 indicates possible problems fitting the PSF.   A\\
& & & value of zero indicates the object was not successfully measured by\\
& & & DAOPHOT, thus there is no photometry nor measured proper motions\\
& & & for this object (columns pmRa, pmRaErr, pmDec, pmDecErr, r, rErr, chi,\\
& & & sharp, and skyMode will all be 0).\\
chi & float32 & & DAOPHOT estimate of the ratio of the observed pixel-to-pixel scatter from the \\
& & & model image divided by the expected pixel-to-pixel scatter from the image \\
& & & profile. \\
sharp & float32 & & DAOPHOT sharpness. \\
skyMode & float32 & mags arcsec$^{-2}$ & DAOPHOT mode of the local sky histogram. \\
flags\tablenotemark{a} & int32 & & SExtractor flags.  See SExtractor documentation for description of bit values. \\
awin\_image & float32 & pixels & SExtractor windowed semi-major axis length. \\
errawin\_image & float32 & pixels & Error in SExtractor windowed semi-major axis length. \\
bwin\_image & float32 & pixels & SExtractor windowed semi-minor axis length. \\
errbwin\_image & float32 & pixels & Error in SExtractor windowed semi-minor axis length. \\
thetawin\_image & float32 & degrees & SExtractor position angle of the semi-major axis.
\enddata
\tablenotetext{a}{The saturation level was incorrectly set for the Bok data,
and thus the saturation bit for that data should be ignored.}
\end{deluxetable*}

\section{Summary}

We have obtained second epoch imaging over 3100 square degrees of sky
within the SDSS footprint, and combined these data with SDSS astrometry to
generate a deep proper motion catalog.  The catalog includes:
\begin{itemize}
\item
1098 square degrees of sky, 95\% complete to $r = 22.0$, based on
observations in good observing conditions with the 90prime camera
on the Steward Observatory Bok 90 inch telescope;
\item
1521 square degrees of sky, 95\% complete to $r = 20.9$, based on
observations in good observing conditions with the Array Camera on the
USNO, Flagstaff Station, 1.3 meter telescope;
\item
and an additional 488 square degrees of sky of lesser quality data,
obtained under poor seeing or partially cloudy skies on both
telescopes.
\end{itemize}
The catalog provides both absolute proper motions (on the system of the
SDSS+USNO-B catalog) and a second epoch of $r$
band photometry.  The statistical errors for the proper motions range
from roughly 5~\masyear\ at the bright end, to 15~\masyear\ at the
survey completeness limits, for a typical epoch difference of 6 years.
The systematic errors for the 1.3m are less than 1~\masyear.  For the
Bok survey, there are FOV-dependent systematic errors of as high as 2
-- 4~\masyear, though typically less than that.  The statistical
errors for the $r$ photometry varies from 0.01 -- 0.02 magnitudes at
the bright end to about 0.07 magnitudes at the completeness limits.

Comparison with the LSPM high proper motion catalog, as well as using
duplicate observations between the Bok and 1.3m surveys, indicates a
completeness rate of 98\% (falling from 99\% at a total proper motion
of 150~\masyear\ to 94\% at a total proper motion of 700~\masyear),
and a contamination rate of false proper motions of 1.5\%.  The
catalog is available in the SDSS CasJobs.

\acknowledgements
We thank Ani Thakar for support loading the catalog
into the SDSS CasJobs.  We also thank Mike Lesser, Edward Olszewski, and
Grant Williams for their outstanding work building 90Prime, and Fred Harris
for his outstanding work building the Array Camera.
MK gratefully acknowledges the support of the
NSF under grant AST-1312678 and NASA under grant NNX14AF65G.  KW
gratefully acknowledges the support of the NSF under grants
AST-0206084 and AST-0602288.  This material is also based on work
supported by the National Science Foundation under grant AST 06-07480.

Funding for the SDSS and SDSS-II has been provided by the Alfred
P. Sloan Foundation, the Participating Institutions, the National
Science Foundation, the U.S. Department of Energy, the National
Aeronautics and Space Administration, the Japanese Monbukagakusho, the
Max Planck Society, and the Higher Education Funding Council for
England. The SDSS Web Site is http://www.sdss.org/.
The SDSS is managed by the Astrophysical Research Consortium for the
Participating Institutions. The Participating Institutions are the
American Museum of Natural History, Astrophysical Institute Potsdam,
University of Basel, University of Cambridge, Case Western Reserve
University, University of Chicago, Drexel University, Fermilab, the
Institute for Advanced Study, the Japan Participation Group, Johns
Hopkins University, the Joint Institute for Nuclear Astrophysics, the
Kavli Institute for Particle Astrophysics and Cosmology, the Korean
Scientist Group, the Chinese Academy of Sciences (LAMOST), Los Alamos
National Laboratory, the Max-Planck-Institute for Astronomy (MPIA),
the Max-Planck-Institute for Astrophysics (MPA), New Mexico State
University, Ohio State University, University of Pittsburgh,
University of Portsmouth, Princeton University, the United States
Naval Observatory, and the University of Washington.

{\it Facilities:} \facility{Bok (90prime)}, \facility{USNO:1.3m (Array Camera)}

\end{document}